# Using math in physics:
# 6. Reading the physics in a graph

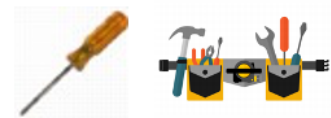

*Edward F. Redish,*
University of Maryland - emeritus, College Park, MD

Learning to use math in physics involves combining ("*blending*") our everyday experiences and the conceptual ideas of physics with symbolic mathematical representations. Graphs are one of the best ways to learn to "build the blend". They are a mathematical representation that builds on visual recognition to create a bridge between words and equations. But students in introductory physics classes often see a graph as an endpoint — a task the teacher asks them to complete — rather than as a tool to help them make sense of a physical system. And most of the graph problems in traditional introductory physics texts simply ask students to extract a number from a graph. But if graphs are used appropriately, they can be a powerful tool in helping students learn to build the blend and develop their physical intuition and ability to think with math.

In this paper, I outline the issues involved in students' learning to use graphs effectively and present a variety of graphical problems to help them build the blend.

This paper is one of a series of papers on using math in physics.[1] Math is used differently in physics than it is presented to students in math classes. Understanding these differences can help instructors understand student struggles and create more effective intuition-building curricula. The series identifies a number of distinct tools, often taken for granted by instructors, that students need to learn in order to "think with math".

In my text and class slides, when I use one of these tools, I note it with a small icon, reminding the students that we are using one of our general collection of methods for building physical knowledge with math. The icon I use for reading the physics in a graph is a screwdriver, shown at the top of this page, since it is one of the basic tools for taking something apart or putting it back together.

Since graphical problems take up a lot of space and my space here is limited, I've put many of the problems I want to share in the Supplementary Materials. I encourage you to explore those as well as the ones given here.

## Graphs are an important component of learning to build the blend

In professional practice, graphs are perhaps THE most important way to code scientific information mathematically. As physicists, we might be prejudiced in favor of powerful sets of equations, such as Maxwell's equations, the Schrödinger and Dirac equations, or the Navier-Stokes equation; and advanced physics texts support this prejudice. But if you look at papers in the top journals that cover multiple sciences, such as *Science* or *Nature,* many of them display information as graphs. Very few use any equations.

But more than that, many students find it hard to get beyond the idea that science is about a list of memorized facts and algorithms (often inferred from their experience in earlier science classes). They find it hard to see math in science as about coding a coherent, sense-making understanding of the physical world and how it works. Graphs can help them learn to see math as sense-making. The reasons behind the success of this approach lie in the basic cognitive science of how we build complex concepts, and these ideas lead us to use graphs in specific blend-directed ways.

I developed many of the problems presented here from my experience as an instructor in classes designed to teach higher order thinking skills in algebra-based physics[2,3] combined with theoretical perspectives drawn from the PER and Cognitive Sciences literatures cited below.

In this paper I focus on using graphs to help students "learn to build the blend of physics and math". I do not discuss the (huge) literature on the role of graphs in the laboratory and in the representation of experimental data.[4]

## Complex concepts begin with physical experience

We create the world we live in from our sensory experience as infants — seeing, touching, moving. Cognitive scientists use the phrase *embodied cognition* to express the idea that we build our understanding of complex and abstract concepts beginning with physical experience[5] via metaphor[6] and blending.[7]

Since vision is the primary tool our brain uses to create the world we perceive,[8] and since graphs are a visual representation, they are an excellent place for students to learn to build the physics-math blend.

### Graphs for the eye and graphs for the mind

Graphs build on the sense of space and location we develop as infants. The most straightforward symbolic representation of that sense is a *map* — a representation of actual physical space — a *graph for the eye*. I present two examples in Figure 1.

On the left is a map with a walk from the College Park Metro Station to the Physics Building at the University of Maryland. If I did that walk and a drone watched me from above, the blue line would be the path I followed — a graph for the eye (or for the camera). On the right of Figure 1 is a 3D map and the track of 3 random walkers diffusing in a gas — a graph for the eye in an important physical situation. Most high school and college students who have used a GPS to get somewhere understand this concept.

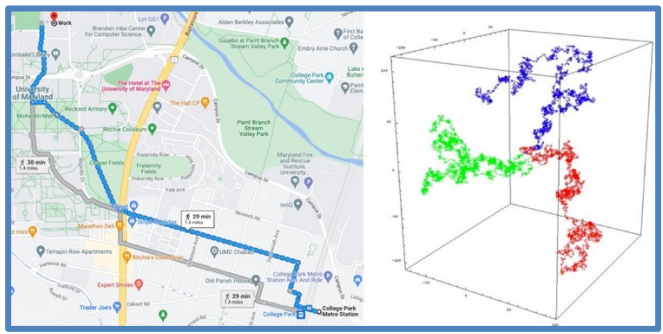

*Figure 1: (L) The track of a walk on a 2D map (Google maps); (R)The track of 3 random walkers on a 3D map*

But most of the graphs that we use in physics are not maps. They extend the concept of map metaphorically, creating *a graph for the mind* — one that needs an analysis or transformation to interpret.[9]

Students have many challenges with learning to use graphs effectively in physics.

The skill of seeing what a graph has to tell you is as challenging to learn as the equation translation skills discussed in previous papers in this series. Learning to use graphs as effective tools in science requires 4 general skills:

- **Constructing** the representation and understanding how it codes mathematical information
- **Interpreting** what the information coded in the graph is telling you about the physical system
- **Connecting** the visual representation in the graph to equations describing the phenomena
- **Knowing when and how** to use graphs appropriately in physical problems.

We need to understand the challenges students face to learning these skills in order to find effective ways to scaffold problems to help them learn to build the physics/math blend.

## Challenges in constructing graphs

Even in a purely mathematical context, constructing a graph requires learning many specific skills including[10]

- creating the palette (axes, coordinates)
- placing and reading points on a graph as associated pairs
- creating associated pairs from a function
- identifying maxima, minima, zeros, crossing points, and other salient features

By the time they've reached an AP physics or college physics class, most students have mastered these skills, at least in the context of creating a graph given a function or a set of pairs. But even in a math context, students may have difficulties that affect their ability to make physical sense with graphs.

## Challenges in interpreting graphs

Student difficulties with reading graphs is the subject of extensive STEM education research.[11,12,13] Here are 4 errors documented in this literature that I have seen from my students in introductory physics classes that create barriers to their building the blend.

- *Treating a graph as a picture* — A student may look at a rising velocity graph and say, "that rise means it's moving up" (or to the right, or anything about where it is in the image rather than a statement about its velocity). This is a "one-step-thinking" quick response that is quite natural,[14] especially if you aren't focused on the difference between "a graph for the eye" and "a graph for the mind".
- *Confusing slope and height* — This is a common conceptual error that we also see when students work with symbolic representations: confusing a value, a change in a value, and a rate of change of that value.
- *Confusing an interval and a point* — This is mostly relevant when students get confused trying to create a derivative at a point rather than realizing an interval is needed.
- Reading positive and negative areas under a curve — When we are looking at a graph for constructing an integral, areas have to be treated as signed quantities. This can be challenging for many students as "area" intuitively seems like "a positive thing".

## Challenges in connecting graphs and equations

A lot of the value of a graph in a physics class is helping students connect graphs to equations. If they learn to blend physical conceptual knowledge with a graph, that can help them bring equations into the blend.

Connecting graphs with equations is a translation skill between 2 very different kinds of symbolic representations. The difficulties students have with these are similar to those in connecting graphs to physical situations; for example, not being able to identify the y-intercept on a graph from an equation or confusing the slope and the value. Many of these are discussed in some of the previously cited references.[11,12,13]

Learning to make these connections can help students understand how equations code for physical meaning in complex situations such as traveling waves or parametric dependences.

## Challenges in knowing when and how to use graphs

But just as with being able to do symbolic math in the context of an abstract math class doesn't imply being able to use those skills in physics, being able to demonstrate graphical skills in math doesn't mean they can do them in physics. Many papers in the PER literature document student difficulties with interpreting the physical information in graphs in a physical context that are analogous to the ones found in math.[15]

Some of the difficulties that students have when using graphs in physics are simply translations of the difficulties seen with any graphs. But some arise because of the additional complexity of using math in physical situations. Because we are asking students to combine knowledge from two distinct domains, abstract math and conceptual physics, many of the difficulties students encounter are *epistemological* — errors about the nature of the knowledge they are learning.

The critical epistemological issue is often *framing*.[16] When you are given a problem to solve, you first must answer the questions, "What's going on here? Given the current situation (problem), what knowledge do I have that I can usefully apply?" This step is often not conscious but can be a response to expectations based on previous experiences in situations perceived as similar. Some framing errors I've often seen are:

- seeing graphs as solutions rather than sense-making tools
- treating the graph as math rather than a way to express physical information
- not seeing that multiple graphs as ways of highlighting different information about a physical situation

If asked to create a graph in a physics problem, students will often just sketch a pair of axes. Often, these will be unlabeled (or marked as "x" and "y" no matter what is actually being plotted) and have no tic marks to indicate scales on either axis.

But if they are to begin to build the blend, they need to frame the task of creating a graph as making a mathematical model of a physical situation, as discussed in paper 4, Toy Models, Figure 1. This requires tying the explicit graph-creation steps given above to physical meaning; asking questions such as, "What does the origin mean physically? What are appropriate units and scales for each axis?" (As Randall Munroe suggests in his xkcd comic, not doing this properly can have consequences outside of losing points in a physics class. See Figure 2.)

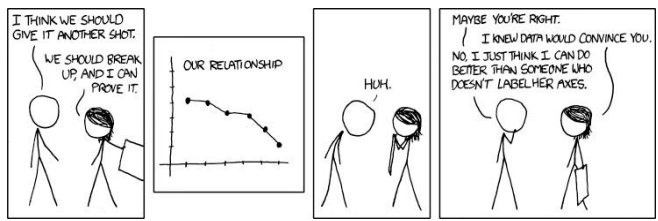

*Figure 2: Labeling your axes is important!*
*Randall Munroe, xkcd, used with permission*

I note that some educational researchers conceive of the difficulty of using math in science as an issue of "transfer" — moving skills learned in one domain into another. In my experience, both as a practicing physicist and as a physics education researcher, this feels misleading. What happens when someone builds good physical intuition with math is not simply a "transfer" of skills. Rather, it is building something new from the joining of distinct mental spaces. Physical concepts are merged with mathematical symbolic representations to create a richer and more powerful mental structure.[17] This is precisely what Fauconnier & Turner[18] describe in their *blending* model of building new concepts that I rely on in this series.

## Carefully designed tasks can help students build the blend

These challenges suggest that, as with equations, it is not enough to assume "They can do it because they learned it in math." To use graphs effectively in physics, students need to do tasks that help them learn to make the appropriate math-physics connections — to read the physics in a graph. This includes a number of diverse skills, including

- interpreting the physics coded in a single graph
- seeing that multiple graphs can highlight different aspects of a physical situation
- checking for consistency among different graphs for a single situation
- seeing graphs as an intermediary between physical concepts and equations

As with any complex skill, a good way to help students is to scaffold by beginning with a simple example and then increasing the level of complexity a step at a time.

### Reading the physics in a graph

*Interpreting the graphs of a single situation*

A good way to start helping students build the blend using graphs is to ask them to interpret the physical correlates of various salient points on a graph for the mind — one that requires some mental processing that asks them to connect the physics and the math. Kinematics offers lots of opportunities for problems of this type. It's a good idea to start with graphs of position vs time since that requires fewer levels of inference. The problem *Dancer position graphs* in the Supplementary Materials is an example. The PhET *Moving Man* simulation[19] is an excellent way to generate such problems.

Once you have worked a few of those in class and for homework, you might try something like the problem shown in Figure 3. I regularly assigned this for homework and, in Course Center hours, I observed it as being highly effective in getting students to struggle appropriately and work out the correct answers together.

A nice next step is to ask about forces in an acceleration graph. An example of this is *The Juggler* in the Supplementary Materials. This problem is *much* more challenging (and interesting to discuss in class) since it requires not only translations between kinematic variables, but also bringing physical principles to bear (Newton's 2$^{nd}$ law and knowledge about the gravitational force).

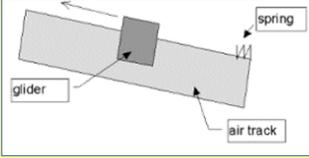

*Figure 3: Interpreting the physics in a single graph*

A more direct but equally challenging example asks students to pick the graph that describes the force/stretch characteristics of *Realistic springs* in the Supplementary Materials.

*Matching multiple graphs to multiple situations*

Since variation is a critical part of learning to make sense of something,[20] a good next step is to vary the physical situation and ask students to choose how the graph changes.

Thornton & Sokoloff introduced one way to do this to great advantage in their ground-breaking concept tests.[21] A number of graphs of a single variable are given and the student's task is to match the graphs to a list of physical situations described in words.

Figure 4 shows one from the FMCE that I consider one of the best prototypes of this type of problem. It's considerably simpler than *The Juggler* problem since it only requires translation from velocity to acceleration and doesn't require integrating concepts about force.

*Learning how multiple graphs provide multiple perspectives*

The next step is to use multiple graphs to describe a single physical situation. This is an outstanding example of the deep epistemological idea that in science we don't just have answers, we have a network of supporting fundamental principles that let us get to an answer in multiple different ways. If you've done position, velocity, and acceleration graphs in your study of kinematics, you've already begun helping your students see this.

To do a problem like this successfully, students have to be able to do the equivalent of "running a mental video of what's happening" and recognize what happens to each variable as the system evolves, a useful skill in developing an understanding of mechanism.[22]

But in physics we may have a dozen different ways of looking at a physical situation! In a situation with motion, we can look at position, velocity, acceleration, multiple kinds of forces, momentum, and energy — total, kinetic, and potential. Figure 5 shows an example that requires students to think about how each different variable says something about a physical situation.

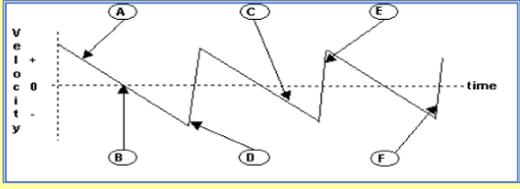

*Figure 4: Matching situations to a set of graphs; Thornton & Sokoloff, FMCE, used with permission*

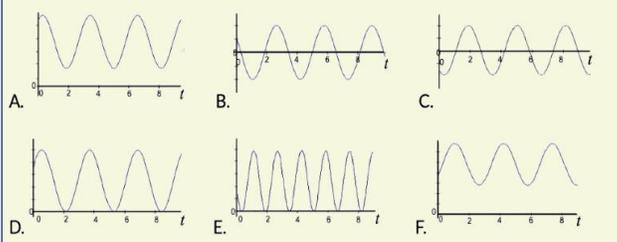

*Figure 5: Matching multiple variables to graphs of a single situation*

I've included some additional examples in the Supplementary Materials.

*Considering multiple situations and multiple graphs*

Finally, we can get seriously challenging, asking students to do everything at once — matching graphs to multiple situations and multiple variables. Since this requires managing many concepts at once, it's a good idea to start with simple graphs. The problem in Figure 6 shows an example where the graphs are all straight lines — but you have to interpret what's happening physically in each case in terms of many variables.

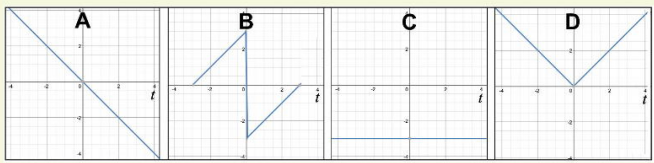

*Figure 6: Matching multiple variables for multiple situations*

## Graphs can help bridge concepts and equations

In math classes, graphs are typically associated with abstract symbolic equations. As we begin to use graphs to help students learn to interpret what's happening physically in a graph, graphs connect what's happening physically and how it's represented in equations.

One of the most important reasons for using symbolic equations rather than putting numbers in right away is that you can explore how a physical phenomenon depends on the parameters in the problem. Online graphing tools such as the Desmos Graphing Calculator[23] can be tremendously useful in this exploration. Problems such as the one in Figure 7 can help students see how valuable it can be to represent physical meaning mathematically, either through graphs or symbolic equations.

We often put physical meaning to a parameter but sometimes to its reciprocal. Here's a problem that starts with a common mathematical form of an equation and bridges to a different and physically relevant parameter.

I give more examples in the Supplementary Materials. A more complex and physical problem (also authentic for life-science students!) is *Opening an ion channel – perhaps*. The problem, *Shifting graphs,* can prepare students to make physical sense of the challenging traveling wave equation, $y(x,t) = A \sin(kx - \omega t)$.

*Figure 7: Using graphs to make sense of parametric dependences*

## Instructional suggestions

Problems of the types shown here are easy to modify slightly so that quiz and exam problems look like class and homework problems but have to be done by using similar reasoning rather than memorizing the answers.

The problems here are presented as multiple choice or matching questions that are easily autograded. Note that in some of these problems the correct choice is "None of these graphs work." Including some of these questions (and having "none" be the right answer) is useful in helping students see that they need to reason through the graph-physics connection in each case and not just focus on the answer.

For each of these types of questions, it's also valuable to have the students generate their own graphs in situations like those described above that can be done in (hand-graded) homework or in-class group work. I've included some examples in the Supplementary Materials with answer keys for the multiple-choice questions. Solutions with explanations are available to instructors in *The Living Physics Portal.*[24]

You can find many more blend-building graph problems in the NEXUS/Physics problem collection in ComPADRE and *The Living Physics Portal.* You can build your own using the marvelous on-line resources available at the University of Colorado's PhET simulations,[25] the Desmos graphing calculator,[23] and ComPADRE.[26]

## Acknowledgements

I would like to thank the members of the UMd PERG over the last two decades for discussion on these issues, and in particular to Chandra Turpen whose questions and suggestions inspired me to write this series. The work has been supported in part by a grant from the Howard Hughes Medical Institute and NSF grants 1504366 and 1624478.

# Supplementary materials for Using math in physics: 6. *Reading the physics in a graph*

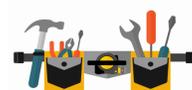

*Edward F. Redish*,
University of Maryland - emeritus, College Park, MD

## Problems in the text and answer keys

For explanations of the solutions, see the *Living Physics Portal*,[1] search *NEXUS/Physics graphs*

---

### Figure 3: Cart on a tilted air track

The graph below shows the velocity graph of a cart moving on an air track as shown in the figure on the right. The track has a spring at one end and has its other end is raised. The cart is started sliding up the track by pressing it against the spring and releasing it. The clock is started just as the cart leaves the spring. Take the direction the cart is moving in initially to be the positive x direction and take the bottom of the spring to be the origin.

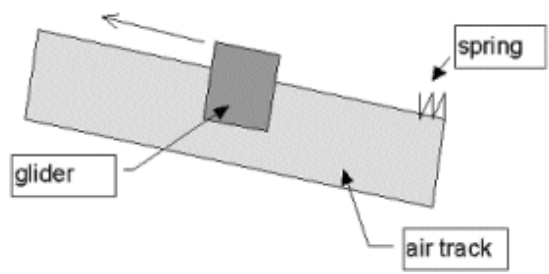

Letters point to six points on the velocity curve. For the physical situations described below, identify which of the letters corresponds to the situation described. You may use each letter more than once, more than one letter may be used for each answer, or none may be appropriate. If none is appropriate, put the letter N.

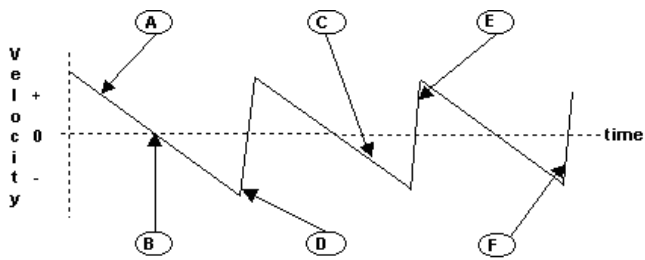

1. This point occurs when the cart is at its highest point on the track.
2. At this point, the cart is instantaneously not moving.
3. This is a point when the cart is in contact with the spring.
4. At this point, the cart is moving down the track toward the origin.
5. At this point, the cart has acceleration of zero.

### Answer key
**1.** B;  **2.** B;  **3.** D, E, F;  **4.** C;  **5.** N

## Figure 4: A moving toy car (from the FMCE)

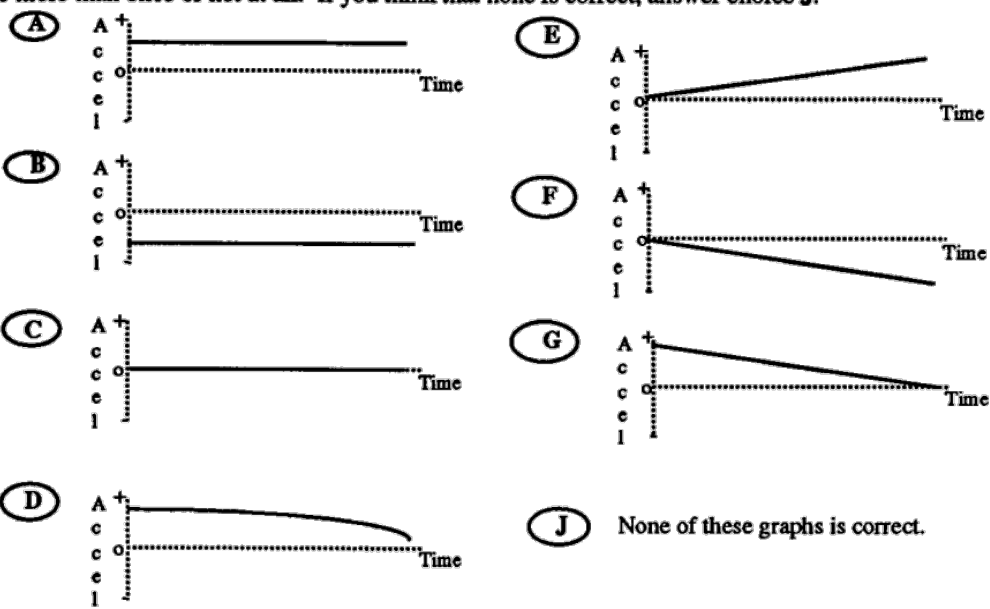

Questions 22-26 refer to a toy car which can move to the right or left along a horizontal line (the + distance axis). The positive direction is to the right.

Different motions of the car are described below. Choose the letter (A to G) of the **acceleration-time** graph which corresponds to the motion of the car described in each statement.

You may use a choice more than once or not at all. If you think that none is correct, answer choice J.

_____ 22. The car moves toward the right (away from the origin), speeding up at a steady rate.
_____ 23. The car moves toward the right, slowing down at a steady rate.
_____ 24. The car moves toward the left (toward the origin) at a constant velocity.
_____ 25. The car moves toward the left, speeding up at a steady rate.
_____ 26. The car moves toward the right at a constant velocity.

### Answer key
**22.** A;  **23.** B;  **24.** C;  **25.** B;  **26.** C

## Figure 5: Oscillating graphs

The position of a mass is hanging from a spring is measured by a sonic ranger sitting 25 cm below its equilibrium position. The mass is started oscillating. Later, the sonic ranger begins to take data.

Below are shown time graphs associated with the motion of the mass.
Graph A shows the mass's position as measured by the ranger.
For each of the following physical quantities, which graph could represent that quantity for this situation? If none are possible, answer N.

1. velocity of the mass
2. net force on the mass

3. force exerted by the spring on the mass
4. kinetic energy of the mass
5. potential energy of the spring
6. gravitational potential energy of the mass

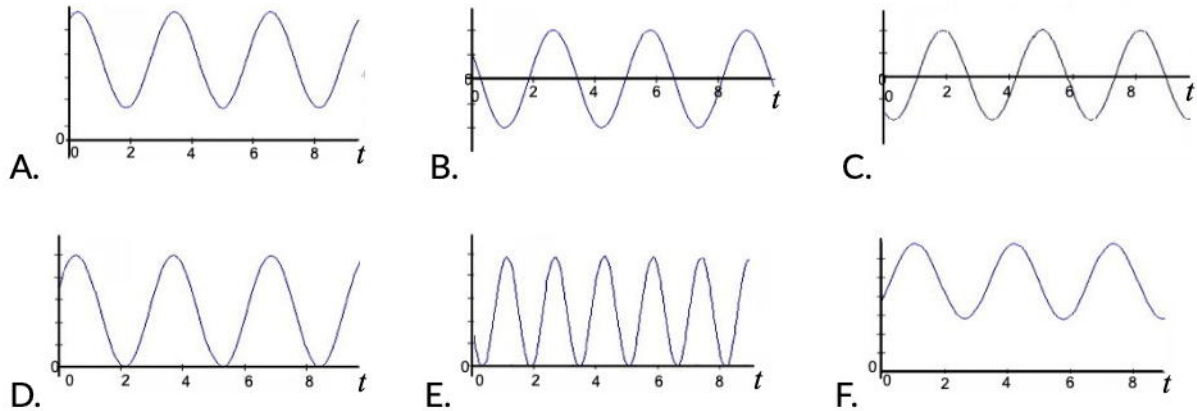

A. B. C.
D. E. F.

*Answer key*

**1.** B;  **2.** C;  **3.** N;  **4.** E;  **5.** N;  **6.** A

## Figure 6: Straigh line graphs of 1D motions

An object's motion is restricted to one dimension along a position axis. In the figure below are shown four possible graphs describing the motion. In each case the horizontal axis represents time, but the vertical axis can represent position, velocity, or acceleration.

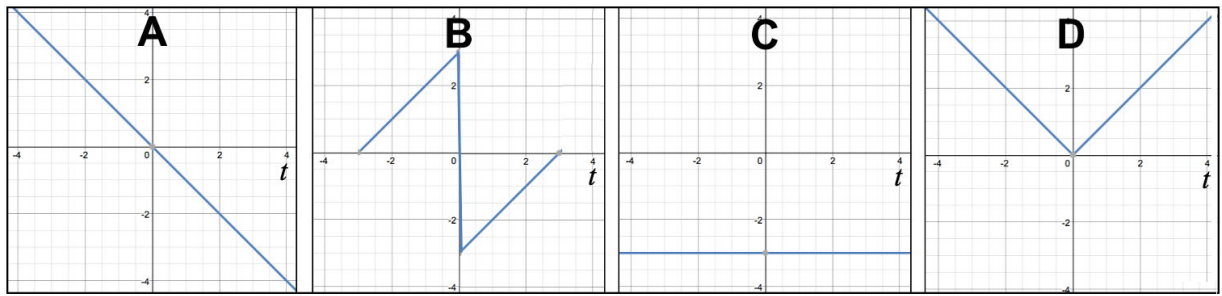

For each of the motions described below in words, select which graph or graphs could describe the motion if the vertical axis were chosen to be an appropriate variable. The variables on the vertical axes could be position, velocity, or acceleration (use the symbols P, V, and A to represent them respectively).

For example, if the motion for one of the items in the list below could be described by graph C if the vertical axis represented velocity, enter C and V in the spaces provided for the graph and variable.

If more than one graph works for a case give both. (If you indicate more than one graph for a case, be clear which graph goes with which variable).

1. A car slowing to stop at a stop sign and then speeding up again.
2. A ball rolling up to a wall and bouncing back.
3. A stationary object.
4. A dropped superball falling on a cement floor.
5. A ball thrown upward.

*Answer key*

**1.** DV;  **2.** DP;  **3.** CP;  **4.** BV;  **5.** AV and CA

---

## Figure 7: Exponential parameters

Explore what changing the parameters, $A$, $b$, and $c$ do to the exponential function using the Desmos graphing calculator. Then answer the questions below.

1. A radioactive element used as a tracer to detect tumors is created in a cyclotron.

If the number of atoms of the element created is $N\_0$, after a time $t$, some of the atoms will have decayed. The total number left at that time will be $N = N_0\, e^{-at}$. It takes 2 hours to deliver the created element from the cyclotron to the medical facility where the test is performed. If you want to choose a tracer that will have more available when the test is performed what value of $a$ should you be looking for?

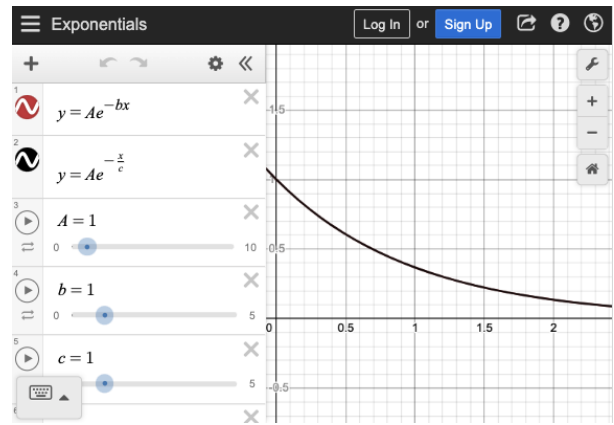

- A. A large value of $a$
- B. A small value of $a$
- C. It doesn't matter

2. The electric potential near a charge in an ionic fluid is given by $V = V_0\, e^{-\frac{r}{\lambda}}$ where $r$ is the distance from the charge and $\lambda$ is the Debye length. At a given distance from the charge, if the Debye length is increased, what happens to the potential?

- A. It increases
- B. It stays the same
- C. It decreases
- D. There is not enough information to decide

*Answer key*

**1.** B;  **2.** A;

# Additional Problems

# Reading the physics in a graph

## Single situation, single variable, single graph

### Dancer position graphs

In a video of a ballet dancer performing a *grand jeté* ("big jump") we can build a "graph for the eye" of her motion by putting a dot on her eye in each frame as shown in the figure.

If instead of just tracking her motion in space we want to create graphs of her position vs time, we have to make sense of how she is moving.

Answer the following questions taking the x-y coordinates describing her motion as shown in the figure with the origin at the point where they cross.

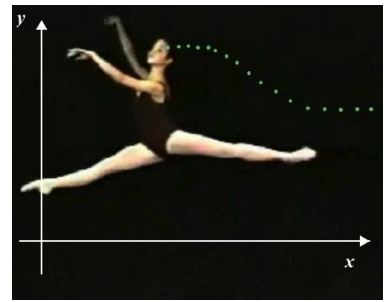

1. Which graph might represent a graph of her x (horizontal) coordinate as a function of time for the time when she is in the air?

2. Which graph might represent a graph of her y (vertical) coordinate as a function of time for the time when she is in the air?

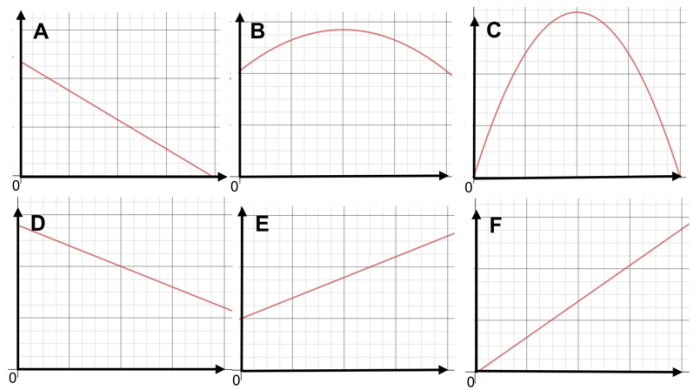

3. Suppose we chose our coordinate system for the ballet dancer's leap so that the origin was on the ground but at the center of the screen rather than at the left. Now which graph might represent a graph of our data of her x (horizontal) coordinate as a function of time?

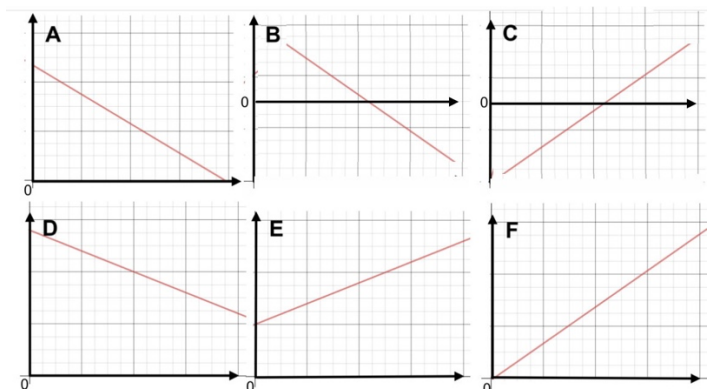

*Answer key*
**1.** A; **2.** B; **3.** B.

*The Juggler: Acceleration*

The picture at the left below shows a single frame from a video of a juggler tossing three balls up and down. In this frame, he has caught the central ball and is holding it so the ball is momentarily at rest. This corresponds to the time t = 0. In the video, he then throws the ball upwards it goes upward, it reaches its highest point a few moments later, and then it comes most of the way down to his hand, but the video stops before he catches it again. Taking position and time data, the video analysis program finds velocity and acceleration for the ball at each time. A smoothed version of the acceleration graph for the range of times described is shown at the right below. (The full video can be watched by clicking on JugglerVideo.)

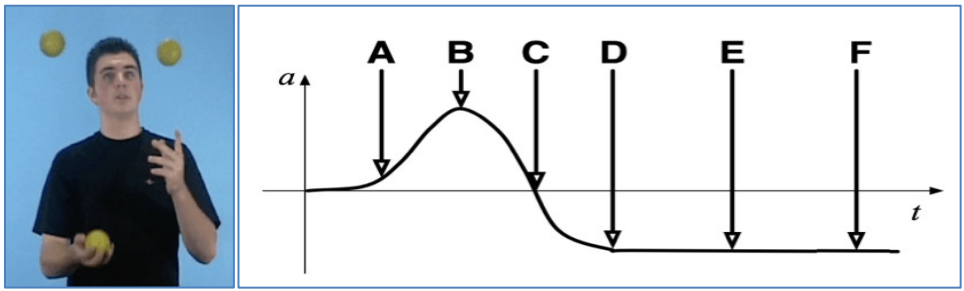

For each of the following physical descriptions of an instant of time, select the letters on the acceleration graph corresponds to those instants. Be sure to give all the instants that most likely match the physical description. If none match, put N.

1. At this instant, the ball is moving upward with its maximum speed.
2. At this instant, his hand releases the ball.
3. At this instant, the ball has velocity 0.
4. At this instant, the ball is traveling upward.

*Answer key*
**1.** C;  **2.** D;  **3.** E;  **4.** A, B, C, D

## The Juggler: Velocity

In the graph at the right is shown the graph of the vertical velocity of a ball being thrown upward by a juggler (figure at the left). Five points are labeled on the graph. (The positive direction of the y axis is up.)

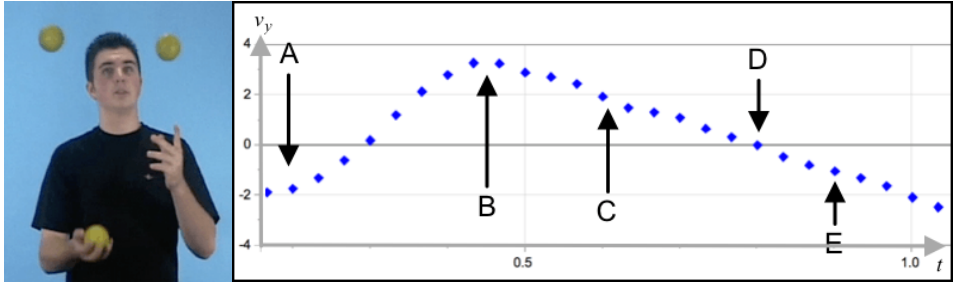

For each of the questions choose all the letters that match the description.
If none match, put N.

1. At which points on the graph is the ball moving upwards?
2. At which point on the graph does the ball have its maximum velocity?
3. At which point on the graph is the ball at its maximum displacement from its starting point?
4. At which point is the ball instantaneously at rest?
5. At which point is the ball accelerating in the upward direction?
6. At which point is the ball accelerating in the downward direction?

## Answer key
**1.** B, C;  **2.** B; **3.** D; **4.** D; **5.** A; **6.** C, D,E

## Hitting a bowling ball

A bowling ball sits on a hard floor at a point which we take to be the origin. The ball is hit some number of times by a hammer. The ball moves along a line back and forth across the floor as a result of the hits. The region to the right of the origin is taken to be positive, but during its motion the ball is at times on both sides of the origin. After the ball has been moving for a while, a sonic ranger is started and takes the following graph of the ball's velocity. (It is positioned on the left of the origin pointing to the right.)

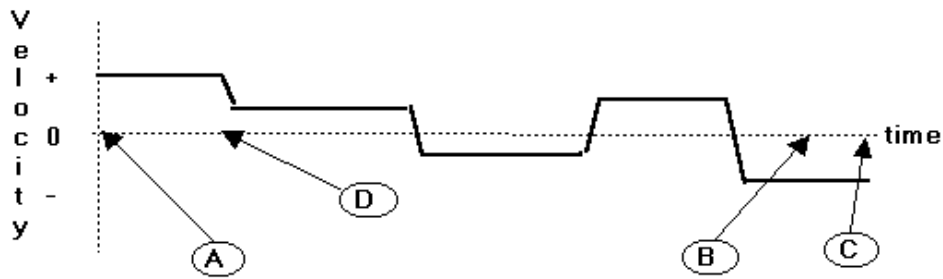

Answer the following questions with the symbols L (left), R (right), N (neither), or C (can't say which). Each question only refers to the time interval displayed by the computer. Explain your reasoning in each case.

1. Which side of the origin is the ball at for the time marked A?
2. At the time marked B, in which direction is the ball moving?
3. Between the times A and C, what is the direction of the ball's displacement?
4. The ball receives a hit at the time marked D. What is the direction the ball is moving after that hit?

## Answer key
**1.** N;  **2.** L;  **3.** R;  **4.** R

## Realistic Springs

Real springs only follow the Hooke's law model for small displacements around their rest length. For a realistic spring like the one shown in the figure at the right, it behaves as follows: for stretching from its rest length, it obeys Hooke's law for a while, then as you stretch it further, it gets stiffer as the coils begin to bend. Eventually it straightens out into a long straight wire which is very hard to stretch at all. If you keep pulling harder, the wire suddenly goes plastic (stretches easily) and breaks. If you try to compress it, the coils almost immediately get pushed together and you can squeeze it very hard without getting much compression.

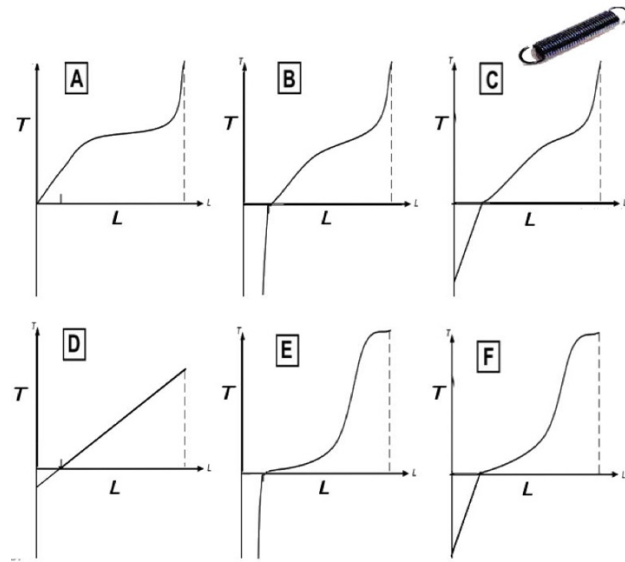

The graphs show the tension (positive for stretch, negative for compression) vs. the length of the spring. Which best represents the Force vs. Length curve for this spring?

## Answer key

E

# Multiple situations, single variable, multiple graphs

## Velocity graphs

An object's motion is restricted to one dimension along the + distance axis. Answer each of the questions below by selecting the velocity graph that is the best choice to describe the answer. You may use a graph more than once or not at all.

1. Which velocity graph shows an object going away from the origin at a steady velocity?
2. Which velocity graph shows an object that is standing still?
3. Which velocity graph shows an object moving toward the origin at a steady velocity?
4. Which velocity graph shows an object changing direction?
5. Which velocity graph shows an object that is steadily increasing its speed?

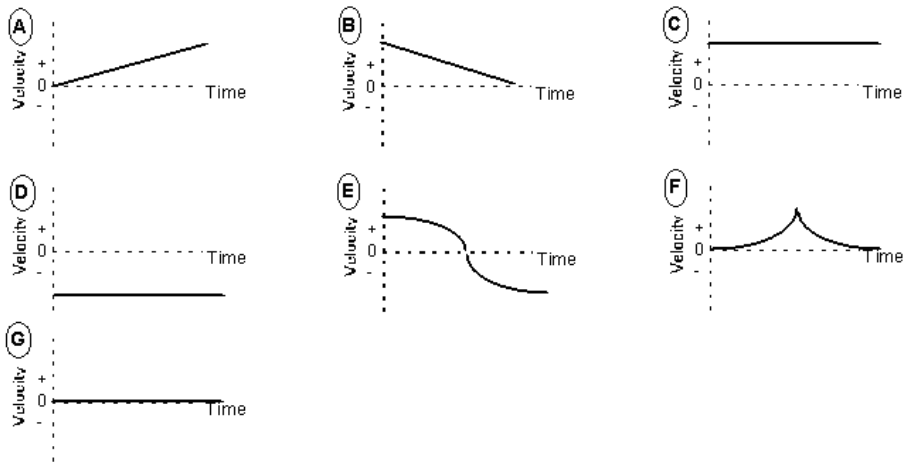

## Answer key
**1.** C; **2.** G; **3.** D; **4.** E.;**5.** A

## Changing double slits[2]

In the figure below are shown graphs of the intensity pattern of light as a function of position on a screen produced by 4 sets of double slits labeled P, Q, R, and S. To produce the graphs, the laser beam and screen are held fixed and the slits simply changed for each other. The graphs are plotted on the same scales for distance (abscissa) and intensity (ordinate).

Rank the relative widths, $d_P, d_Q, d_R, d_S$, and the separations of the slits, $a_P, a_Q, a_R, a_S$.

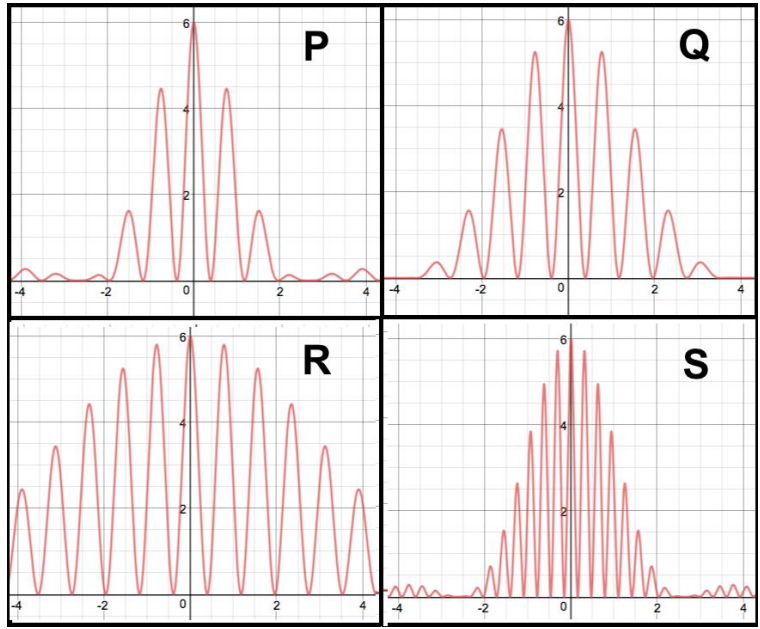

## Answer key

**1.** $d_P = d_S > d_Q > d_R$

**2.** $a_S > a_P = a_Q = a_R$

## Single situations, multiple variables, multiple graph

These problems are particularly good for focusing on consistency between graphical descriptions of a situation representing different variables. I encourage you to grade these with partial credit for consistency. That is, if a student can demonstrate that they started with one wrong graph, but the others follow correctly from those assumptions, I would give credit for the correct reasoning.

Doing this also sends students the message that what matters is learning the reasoning rather than memorizing the answers.

### Kinetic energy and momentum graphs

The figure at the right shows two pucks that can slide on a frictionless table. Puck II is four times as massive as puck I. Starting from rest, the pucks are pushed across the table by two equal forces. Below are shown 8 possible graphs. Each graph displays some property of the two pucks as a function of time.

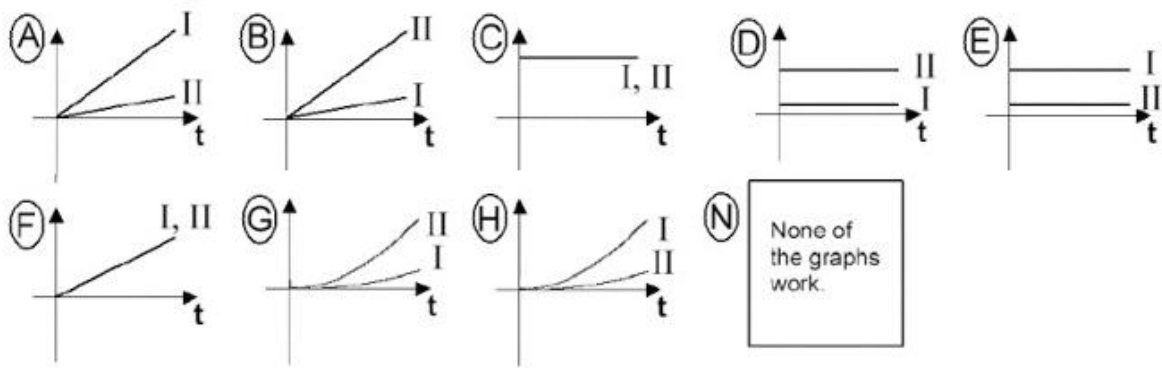

1. If the graphs represent **speed** which graph would correctly represent the speeds of the two pucks?
2. If the graphs represent **momentum** which graph would correctly represent the momenta of the two pucks?
3. If the graphs represent **kinetic energy** which graph would correctly represent the kinetic energy of the two pucks?
4. If they start at the same time, which puck will cross the finish line first?

### Answer key
**1.** A; **2.** F; **3.** H; **4.** I

## The Moving Man simulation

The Moving Man (from the PhET simulation of the same name) walks steadily towards the house for 6 seconds, then stands still for 6 seconds, and then towards the tree at the same speed as before for 6 seconds. Which graph would the program show for his position (*x*)? Pick either A, B, C, or D (or N). Which graph would the program show for his velocity (*v*)? Pick either E, F, G, or H (or N).

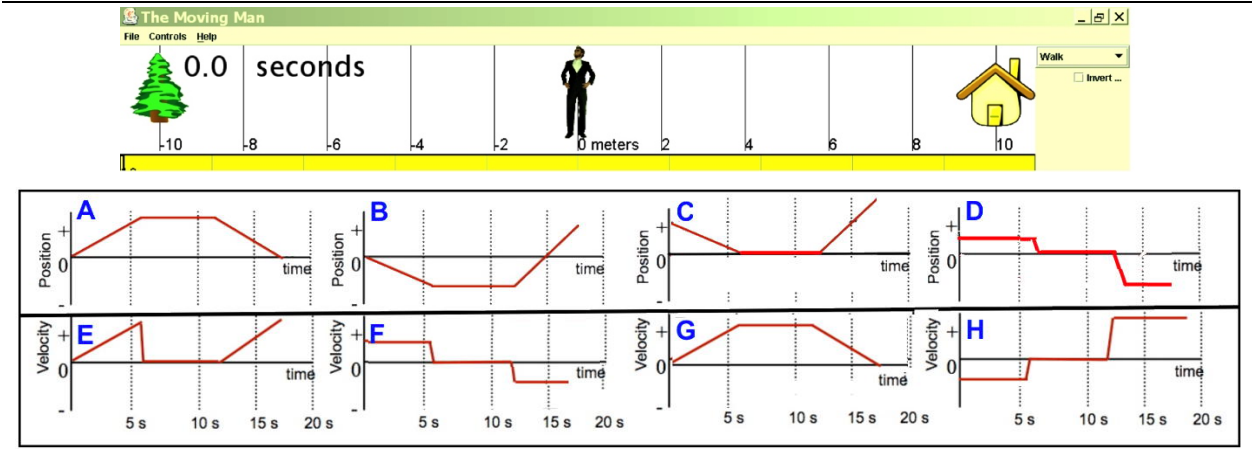

## Answer key

**x:** A
**v:** F

## The TA on a skateboard

In the picture shown at the right (a single frame from a video), the TA shown is riding on a skateboard that is moving to the left with a constant velocity. Just before this frame, he threw the ball straight up. Imagine taking data from this video by placing a dot on the center of the ball from the time just after it left his hand until the ball crosses the y-axis.

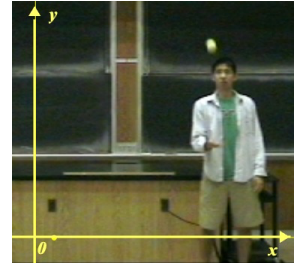

Suppose you made time graphs of the x and y position, velocity, and acceleration, and then fit each of your graphs with smooth curves. Which of the graphs in the figure below would look like each of your graphs if the vertical axis were given the right units and scale? If none could work write N.

| For the x-coordinate | | For the y-coordinate | |
|---|---|---|---|
| Postion | | Postion | |
| Velocity | | Velocity | |
| Acceleration | | Acceleration | |

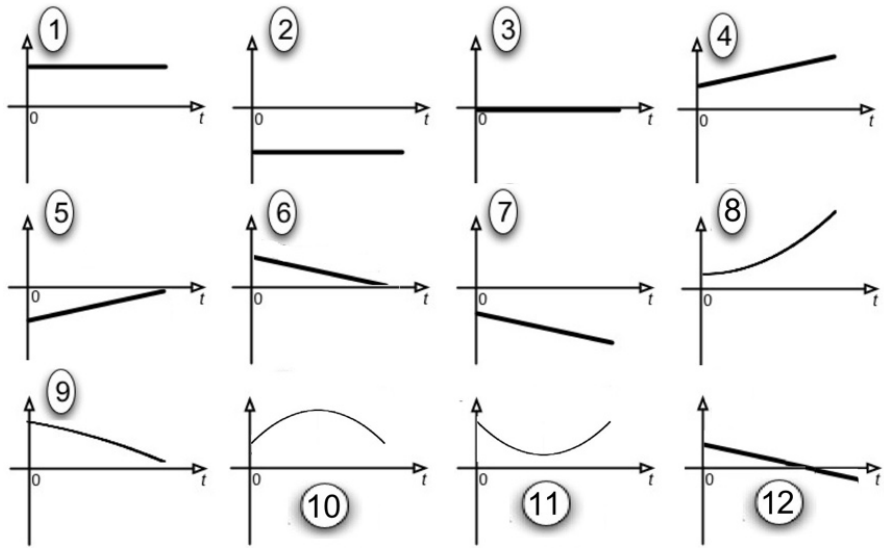

## Answer key
x: P 4; V 2; A 3
y: P 10; V 12; A 2

## The perfect bouncer

A superball is dropped from a height of 1 m and bounces a number of times before it is caught. Below are shown graphs of some of the physical variables of the problem.

Match the graphs that best show the time dependence of the variables in the list below the graphs. (Assume for these first few bounces the superball can be treated as a "perfect bouncer.")

You may use a graph more than once or not at all. If none of the graphs work well for a variable, put N. Note: the time axes are to the same scale, but the ordinates {"y axes"} are not.

The time $t = 0$ is taken to be the instant when the ball leaves the hand. Use a coordinate system in which the positive direction is taken as up and the origin is at the floor.

1. The velocity of the ball
2. The kinetic energy of the ball
3. The potential energy of the ball
4. The momentum of the ball
5. The total mechanical energy of the ball.
6. The position (y coordinate) of the ball

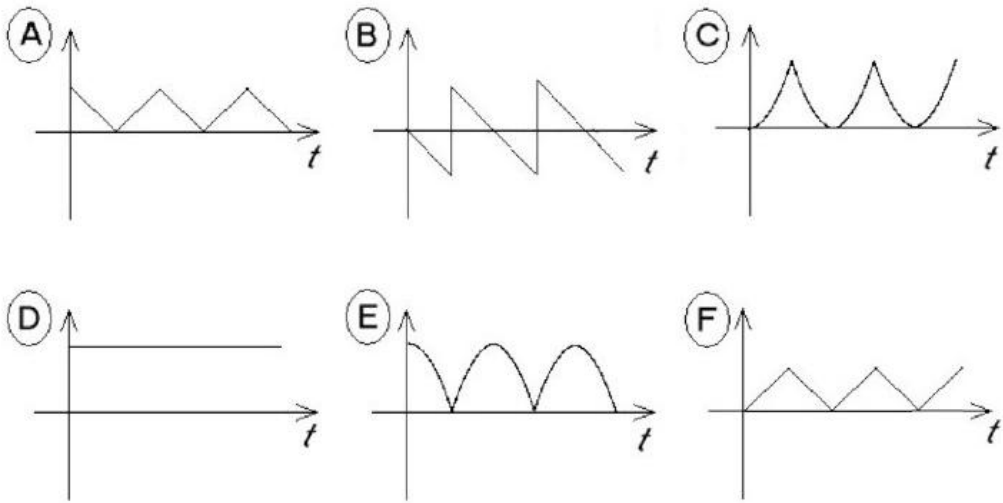

## Answer key

**1**. B; **2**: C; **3**. E; **4**: B; **5**: D; **6**. E

## Tracking round a circuit

The circuit shown in the diagram at the right contains a battery and 3 resistors. The battery has an EMF of 5V, $R_1=2\ \Omega$, $R_2=3\ \Omega$, and $R_3=5\ \Omega$.

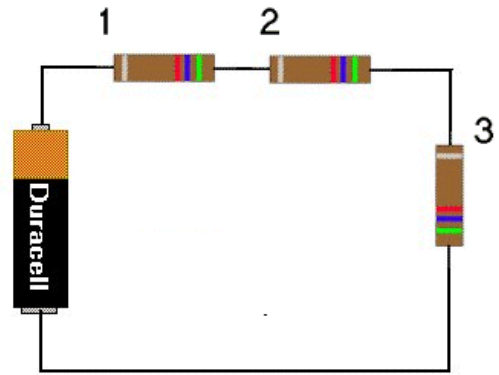

Below are shown 3 graphs tracking some quantity around the circuit. On the first, plot the voltage a test charge would experience as it moved throughout the circuit. On the second, plot the electric field a test charge would experience as it moved through the circuit. On the third, plot the current one would measure crossing a plane perpendicular to the wire of the circuit as one goes through the circuit.

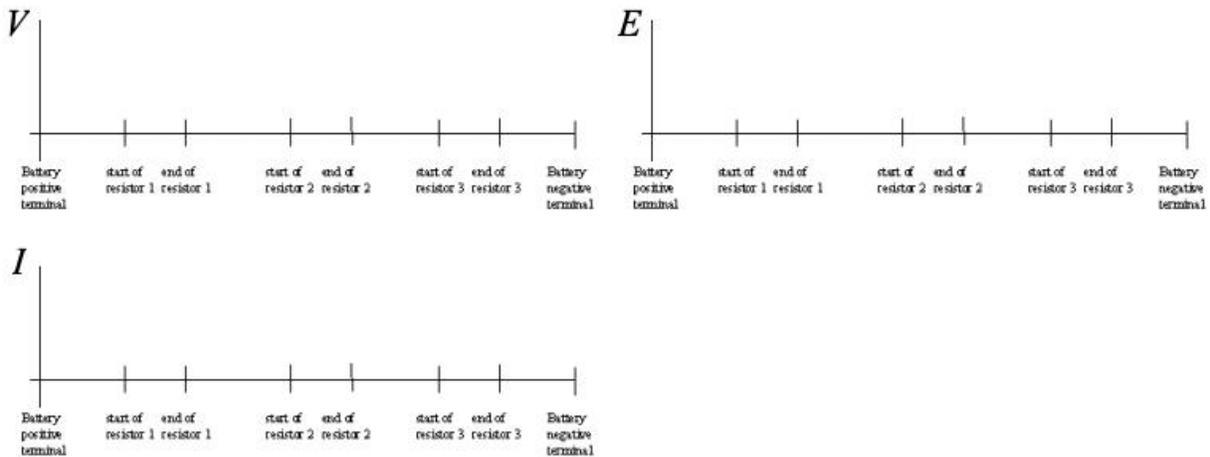

## Answer key

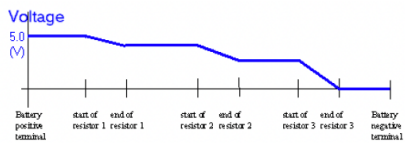

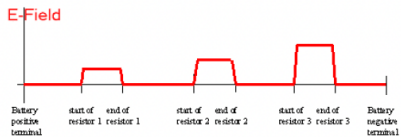

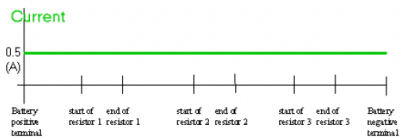

## Springy cart on a track

A cart with a spring on its end is moving to the right on a frictionless air track. It keeps going until it hits a wall, the spring compresses, and the cart bounces off the wall and moves back in the opposite direction.

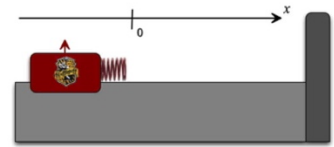

Choose all the graphs that could represent each of the following quantities as a function of time.

1. Position of the cart
2. Velocity of the cart
3. Momentum of the cart
4. Kinetic energy of the cart
5. Potential energy of the spring
6. Total mechanical energy of the cart and spring

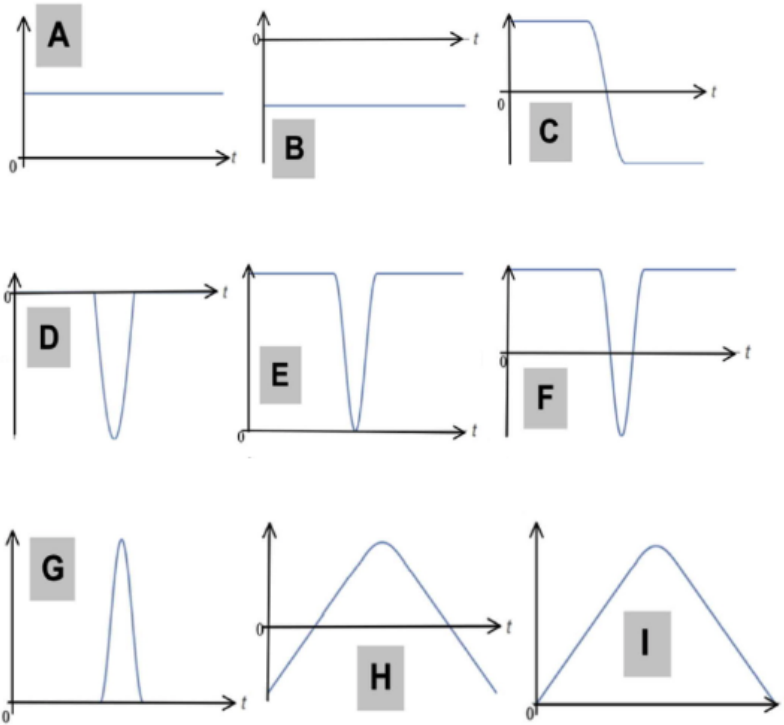

## Answer key
**1.** H; **2.** C; **3.** C; **4.** E; **5.** G; **6.** A

## Pushing a carriage

A young father is pushing a baby carriage at a constant velocity along a level street. A friend comes by to chat and he lets go of the carriage. It rolls on for a bit, slows, and comes to a stop. On the graphs below, sketch qualitatively accurate (i.e., we don't care about the values but we do about the shape) graphs of each of the indicated variables. Take the positive direction as the direction he is walking.

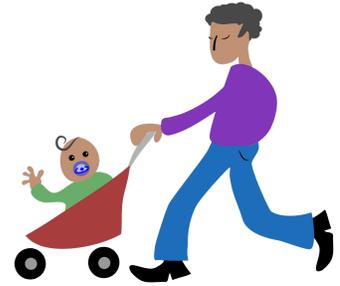

At time $t = 0$ he is moving with a constant velocity. At time $t_1$ he releases the carriage. At time $t_2$ the carriage comes to rest.

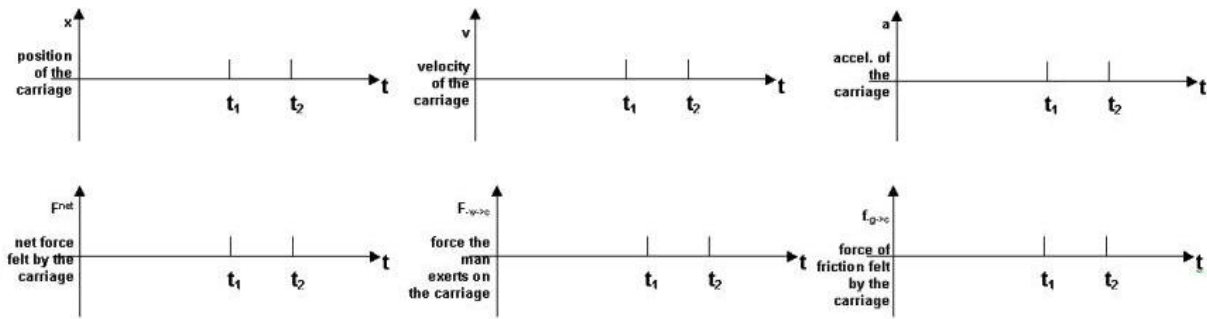

## Answer key

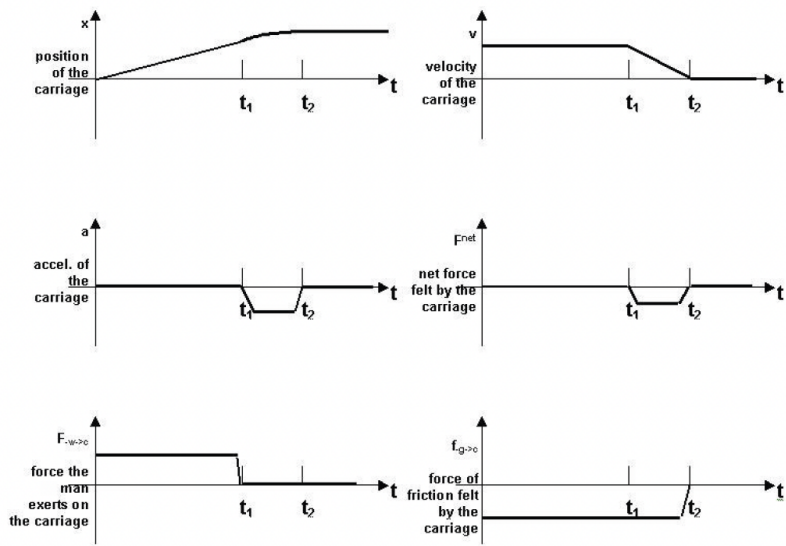

## Graphing a pulse on a string

Consider the motion of a pulse on a long taut string. We will choose our coordinate system so that when the string is at rest, the string lies along the x axis of the coordinate system. We will take the positive direction of the x axis to be to the right on this page and the positive direction of the y axis to be up. Ignore gravity. A pulse is started on the string moving to the right. At a time $t_0$ a photograph of the string would look like figure A below. A point on the string to the right of the pulse is marked by a spot of paint.

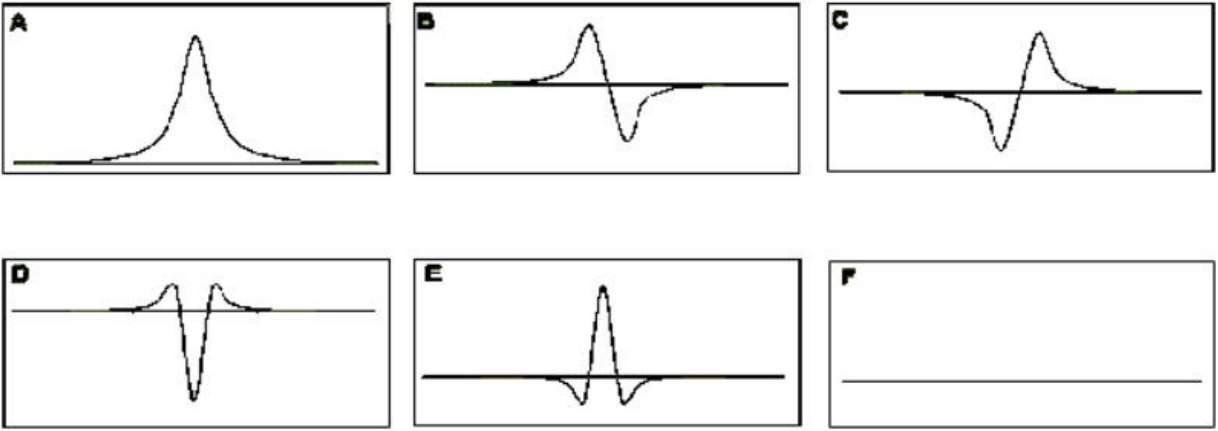

For each of the items below, identify which figure above would look most like the graph of the indicated quantity as a function of position along the string. (Take the positive axis as up in each chase.) If none of the figures look like you expect the graph to look, write N.

1. The graph of the y displacement of the spot of paint as a function of time.
2. The graph of the x velocity of the spot of paint as a function of time.
3. The graph of the y velocity of the spot of paint as a function of time.
4. The graph of the y component of the force on the piece of string marked by the paint as a function of time.

## Answer key
1. A; 2. F; 3. B; 4. D

# Multiple situations, multiple variables, multiple graphs

*Straight line graphs of 1D motions*

## SHO Energies

A mass is connected to a spring and is hanging down at rest as shown in the figure at the right. The dotted line indicated where the spring would be at rest if no mass were connected to it. For the period of time we are considering (2.5 s2.5 s), internal damping of the spring and air resistance can be ignored. We consider two cases:

A- We pull the mass down by 10 cm10 cm and release it.
B- We pull the mass down by 30 cm30 cm and release it.

In the figures below are shown graphs of the y-position vs. time using the ruler as scale (but in unit of meters rather than cm), labeled "A" and "B" for our two cases.

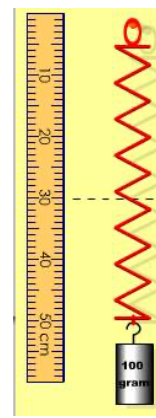

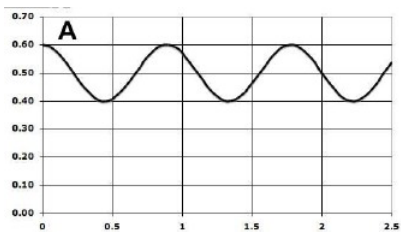
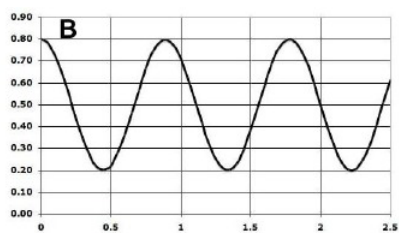

The eight graphs at the right were calculated on a spreadsheet and represent the gravitational PE, the elastic PE, the KE, and the total energy for each of the two cases. The units are Joules. Fill out the table by matching the graphs below with each of these quantities in these cases A and B. If none work, put 9.

| Answer key | Case A | Case B |
|---|---|---|
| **Gravitational PE** | 5 | 6 |
| **Elastic PE** | 1 | 8 |
| **Kinetic Energy** | 3 | 4 |
| **Total Energy** | 7 | 2 |

# Graphs as a bridge:
# Connecting graphs and equations

## Shifting graphs

If the graph labeled A represents a function $f(x)$ match the list of functions on the right with the choices on the left,

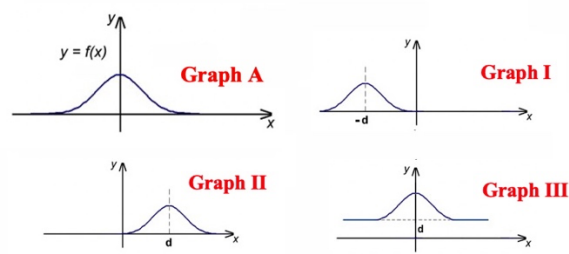

| 1 | $y = f(x + d)$ | A | Graph I |
|---|---|---|---|
| 2 | $y = f(x - d)$ | B | Graph II |
| 3 | $y = f(x) + d$ | C | Graph III |
| 4 | $y = f(x) - d$ | D | None of these |

*Answer key*
**1.** A; **2.** B; **3.** C; **4.** D

## Graphs can support authenticity

Physics majors tend to "cut us some slack", in that they generally accept that what they are being taught is probably relevant for the career they have chosen.[3] But other populations (most of our students) are not necessarily so accommodating.

Engineering students sometimes see the "toy models"[4] that dominate a physics class as having little relevance to their careers as "real-world engineers".[5] You can use graphs as part of problems that imbed the toy model in a more realistic treatment to help them both appreciate models and make the connection to realistic situations.

Pre-health-care professionals in an introductory physics class are one of our most challenging populations. They often do not see the skills we are helping them learn as ones that will be useful in their later careers.[6] We need to include problems that they perceive of as providing insights into rules they are given to memorize in other science classes. These can help them feel that physics has authentic value. I've included a number of examples using graphs in the Supplementary Materials. For more, see *Problems with biological / medical relevance* on the NEXUS/Physics problem collection page.[7]

Here's a problem that serves as a connection between graphs and equations, a connection between equations and the physical situation, and can be seen as authentic by life-science students!

# Opening an ion channel – perhaps

Many biological systems are well-described by the laws of statistical physics. A simple yet often powerful approach is to think of a system as having only *two* states. For example, an ion channel may be open or closed. In this problem, consider a simple model of membrane channels for ions: The system is described by a Boltzmann distribution with only two states, with energies $\epsilon_1$ (open) and $\epsilon_2$ (closed). Assume the "open" state is the state of higher energy, so that $\epsilon_1 > \epsilon_2$.

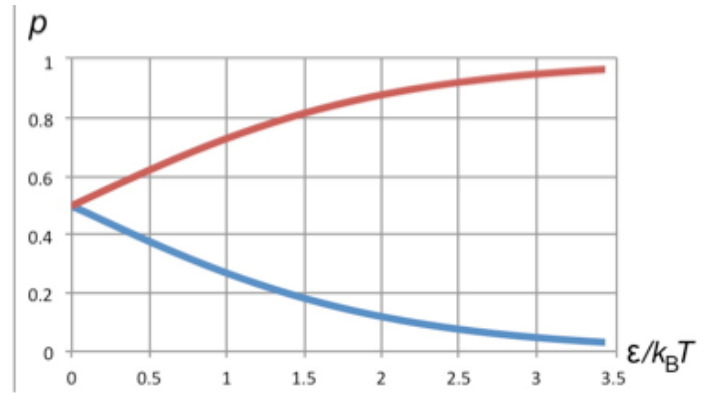

1. If the probability of finding an ion channel open is $P_{open}$ and the probability of finding the ion channel closed is $P_{closed}$, which of the expressions below best represents the relative probability of open to closed, $R = P_{open}/P_{closed}$?

   A. $e^{-\frac{\epsilon_1}{k_BT}} - e^{-\frac{\epsilon_2}{k_BT}}$
   B. $e^{-\frac{\epsilon_2}{k_BT}} - e^{-\frac{\epsilon_1}{k_BT}}$
   C. $e^{-\frac{\epsilon_1}{k_BT}}/e^{-\frac{\epsilon_2}{k_BT}}$
   D. $e^{-\frac{\epsilon_2}{k_BT}}/e^{-\frac{\epsilon_1}{k_BT}}$
   E. Something else

2. Since channels are either open or closed, it must be the case that $P_{open} + P_{closed} = 1$. You can therefore express the absolute probability of finding the channel open as $P_{open}/(P_{open} + P_{closed})$ and work out an explicit expression for it.
Which of the expressions below best represents the absolute probability of finding the channel open?

   A. 0.50.5
   B. $e^{-\frac{\epsilon_1}{k_BT}}$
   C. $1/(e^{-\frac{\epsilon_1}{k_BT}} + e^{-\frac{\epsilon_2}{k_BT}})$
   D. $e^{-\frac{\epsilon_1}{k_BT}}/(e^{-\frac{\epsilon_1}{k_BT}} + e^{-\frac{\epsilon_2}{k_BT}})$
   E. None of the above

3. The two curves in the figure to the right show the probabilities of finding an ion channel either open or closed as a function of $\frac{\epsilon}{k_BT} = \frac{\epsilon_1-\epsilon_2}{k_BT}$. Think about the way accessible energy states, probability, and temperature work together, examine the figure, and identify the true statements in this list:

   A. The top curve is the probability of finding the ion channel open.
   B. The top curve is the probability of finding the ion channel closed.

C. The curves show that the probability of finding the ion channel open is greater at low temperature.
D. The curves show that the probability of finding the ion channel open is greater at high temperature.
E. If $(\epsilon_1 - \epsilon_2)$ is much less than $k_B T$, the relative probability $R$ is approximately 1.

4.1 At approximately what value of $\epsilon/k_B T$ are the ion channels three times more likely to be closed than open?
4.2 If a membrane could change its value of $\epsilon$, what change would it make to open more ion channels?

*Answer key*
1. C; 2. D; 3. B; 4.1 D; 4.2 E

---

## Random or not, here I come

In his physics lab, Radagast has observed the motion of an E. coli bacterium using a video camera. His log-log plot of the square deviation of his chosen bacterium as a function of time is shown in the figure below. (The graph is presented in two equivalent forms: one for those who have used Excel – on the left – instead of log-log graph paper – on the right.) The bacterium seems to have two distinct behaviors: for times shorter than 1 second (marked A) and for times longer than 10 seconds (marked B).

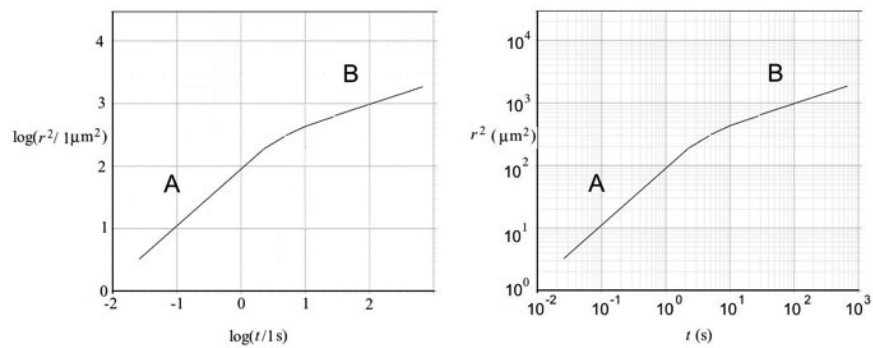

Here are some hypotheses for what might be causing the two different behaviors. Which do you think might be appropriate in region A? In region B?
A. The bacterium is moving purposefully in response to some chemical gradient.
B. The bacterium is moving at random in response to the thermal motion of its environment.
C. The bacterium is constrained in some way.
D. The bacterium is using its flagella (like propellers) to move at a constant velocity
E. The bacterium is accelerating in response to a force in a fixed direction.
F. None of these behaviors are consistent with that part of the graph.

*Answer key*
**In region A:** B: **In region B:** C